\documentclass[a4paper,UKenglish,cleveref]{lipics-v2021}

\title{Source Assignment: Hardness, Approximation, and Parameterized Algorithms}
\titlerunning{Source Assignment: Hardness, Approximation, and Parameterized Algorithms}

\author{Ruxandra Marinescu-Ghemeci}{Faculty of Mathematics and Computer Science, University of Bucharest, Romania}{verman@fmi.unibuc.ro}{}{}
\author{Adrian Micl\u au\c s}{Faculty of Mathematics and Computer Science, University of Bucharest, Romania}{adrian.miclaus1@gmail.com}{}{}
\author{Alexandru Popa}{Faculty of Mathematics and Computer Science, University of Bucharest, Romania}{alexandru.popa@fmi.unibuc.ro}{}{} 

\authorrunning{R. Marinescu et al.}

\Copyright{Ruxandra Marinescu, Adrian Micl\u au\c s, and Alexandru Popa}



\usepackage{amsmath}
\usepackage{amssymb}
\usepackage{mathtools}
\usepackage{microtype}

\nolinenumbers

\newtheorem{problem}{Problem}

\newcommand{\ESA}{\textsc{Explicit Source Assignment}}
\newcommand{\CSA}{\textsc{Complete Source Assignment}}
\newcommand{\SCP}{\textsc{Set Cover with Pairs}}
\newcommand{\MWMCSP}{\textsc{Minimum-Weight Multicolored Subgraph}}

\newcommand{\Occ}{\operatorname{Occ}}
\newcommand{\OPT}{\operatorname{OPT}}
\newcommand{\poly}{\operatorname{poly}}
\newcommand{\bd}{\partial}

\begin{document}
\maketitle

\begin{abstract}
In Source Assignment, each pointer represents a substring of a common reference string and may have several possible source occurrences. The goal is to choose one source for each pointer while minimizing the number of distinct reference boundaries used by the selected sources. The problem was introduced by Qiu and Kingsford (Bioinformatics 2021) in the context of compact genome-graph construction. We distinguish \emph{Explicit Source Assignment}, where admissible source sets are given explicitly, from \emph{Complete Source Assignment}, where all occurrences of each pointer string are available.

We show that Explicit Source Assignment has an exact value-preserving formulation as cardinality Set Cover with Pairs. For every fixed $\varepsilon>0$, it cannot be approximated within $(1-\varepsilon)\ln p$ unless $\mathrm{P}=\mathrm{NP}$, even over a binary alphabet when all pointers have length one, while known algorithms for Minimum-Weight Multicolored Subgraph give an $O(\sqrt{q\log p})$-approximation for arbitrary pointer lengths. Our main hardness result proves that Complete Source Assignment is APX-hard even over a binary alphabet, with equal pointer lengths and at most two sources per pointer. For edge-disjoint equal-length instances, we give a $3/2$-approximation and a $1.12585$-approximation when each pointer has at most two sources. We also introduce the rainbow-path length $\lambda$: the problem is polynomial-time solvable for $\lambda\le2$, and for $\lambda\le3$ we obtain a $(9/8+\varepsilon)$-approximation.

We further give a polynomial-time test for deciding whether abstract source sets can be realized exactly as complete occurrence sets over an unrestricted alphabet. Finally, we give exact fixed-parameter algorithms for several structural parameters, and show that Complete Source Assignment is fixed-parameter tractable in the alphabet size and maximum pointer length.
\end{abstract}

\section{Introduction}

\subsection{Motivation}

Many individuals of the same species share most of their DNA, but they also differ through single-nucleotide variants, insertions, deletions, and larger structural changes. A single linear reference genome cannot represent all of this variation equally well. Genome graphs address this problem by storing several possible sequence alternatives in one structure, which can better represent the diversity present in a population~\cite{PatenEtAl2017}. Such graphs can be used for tasks such as read mapping, variant analysis, and indexing large collections of related genomes.

As more genomes are added, however, the graph can become large. Its size affects memory use, storage, and the cost of later computations. It is therefore useful to build a graph that represents the same genomic sequences while using as few graph nodes and boundaries as possible.
Qiu and Kingsford~\cite{QiuKingsford2021} connected this problem with string compression through the external pointer macro model. Instead of storing every repeated sequence explicitly, a genome can refer to a substring of a common reference. The same substring may occur at several positions in the reference, so one pointer can have several possible sources. All of these sources represent the same sequence, but different source choices can create different boundaries in the resulting genome graph. Choosing the sources carefully can therefore produce a smaller graph without changing the represented genomes.

This led Qiu and Kingsford to introduce the Source Assignment problem: choose one source for each pointer so that the total number of distinct reference boundaries is minimized. They gave an integer linear programming formulation and a preprocessing rule for isolated sources, but left the computational complexity of the problem open~\cite{QiuKingsford2021}. Understanding this complexity is important because exact optimization may become expensive when pointers have many possible sources. This motivates studying hardness, approximation algorithms, and exact algorithms that are efficient when natural structural parameters are small.

There is also a difference between the formal optimization problem and the natural compression setting. In the formal problem, the admissible source sets are given explicitly. In a reference-based compression setting, however, every occurrence of the corresponding pointer string may naturally be considered a possible source. We therefore study both the original explicit-source model and the more restricted complete-occurrence model.

\subsection{Informal problem definition}

Consider a reference string $R$. A pointer $P_i$ represents a substring of $R$. The same substring may appear several times in $R$, and each such occurrence is a possible \emph{source} for the pointer. If a source starts at position $a$ and the pointer has length $|P_i|$, then the source uses the two reference boundaries $a$ and $a+|P_i|$. A source assignment chooses one source for every pointer and aims to minimize the total number of distinct boundaries used by the chosen sources.

In \ESA{}, each pointer is given a nonempty set $S_i$ of allowed sources. These sources must correspond to valid occurrences of the pointer string in $R$, but not all occurrences need to be included. In \CSA{}, the source set is fixed by the reference: every occurrence of the pointer string in $R$ must be included as a possible source. 
This difference is important. In \ESA{}, two pointers representing the same string may still have different source sets. In \CSA{}, equal pointer strings have the same set of occurrences and can therefore be merged. After merging them, sources can be viewed as edges between reference boundaries, with pointer identities as edge colors. Unlike an arbitrary edge-colored graph, however, the edges of each color must be exactly the occurrences of one substring in the same reference string.

\subsection{Previous and related work}

Macro-based compression and textual substitution were studied by Storer and Szymanski~\cite{StorerSzymanski1982}. Relative Lempel--Ziv compression was later used for collections of highly similar genomes~\cite{KuruppuPuglisiZobel2010}. Qiu and Kingsford~\cite{QiuKingsford2021} connected these compression ideas with genome-graph construction. They showed how to transform between genome graphs and the external pointer macro model, introduced Source Assignment, and proposed RLZ-Graph, a genome graph based on Relative Lempel--Ziv compression.

For Source Assignment, Qiu and Kingsford gave an integer linear programming formulation that finds an optimal assignment. Since short pointers may have many possible sources, they also introduced a preprocessing rule that removes isolated sources without changing the optimum. In addition, they considered simple practical choices such as the leftmost and lexicographically smallest source. Their experiments showed that optimizing the source assignment can reduce the number of nodes in the resulting genome graph. However, they left the computational complexity of Source Assignment open~\cite{QiuKingsford2021}. Our hardness results answer this question, while our approximation and parameterized algorithms give alternatives when solving the ILP exactly is too expensive.

The explicit-source problem is closely related to classical covering problems. In Set Cover with Pairs~\cite{HassinSegev2005}, an element is covered when both objects of one of its allowed pairs are selected. In \ESA{}, the objects are reference boundaries, and the two endpoints of a source form such a pair. The same problem can also be viewed as a unit-weight Minimum-Weight Multicolored Subgraph instance~\cite{HajiaghayiEtAl2006}.

When every edge has only one color, this graph problem becomes Minimum Rainbow Subgraph. Its complexity and approximation properties have been studied in several works~\cite{KatrenicSchiermeyer2011,HueffnerEtAl2015,Schiermeyer2013}. In particular, H\"uffner et al.~\cite{HueffnerEtAl2015} proved APX-hardness even on properly edge-colored paths in which every color appears at most twice. We use this restricted case in our hardness reduction. For our approximation results, we use the matching approach of Camacho, Schiermeyer, and Tuza~\cite{CamachoSchiermeyerTuza2010}, the MAX 2-AND approximation of Brakensiek et al.~\cite{BrakensiekEtAl2026}, and the $(4/3+\varepsilon)$-approximation for 2--3-Set Packing of Neuwohner~\cite{Neuwohner2023}.

\subsection{Our results}

We first settle the complexity question for the explicit-source formulation. We show that \ESA{} has an exact value-preserving formulation as cardinality Set Cover with Pairs and as unit-weight Minimum-Weight Multicolored Subgraph. A direct approximation-preserving reduction from Set Cover shows that, for every fixed $\varepsilon>0$, \ESA{} cannot be approximated within $(1-\varepsilon)\ln p$, unless $\mathrm{P}=\mathrm{NP}$, even over a binary alphabet when all pointers have length one. On the positive side, known algorithms for Minimum-Weight Multicolored Subgraph give an $O(\sqrt{q\log p})$-approximation for \ESA{} with arbitrary pointer lengths.

Our main hardness result concerns the more restricted complete-occurrence model. We prove that \CSA{} is APX-hard even over a binary alphabet when all pointers have the same length and every pointer has at most two sources. The reduction starts from properly edge-colored paths and uses a synchronizing binary encoding that preserves complete occurrences, source multiplicities, and shared boundaries. The constructed instances are also unchanged by the isolated-source pruning rule of Qiu and Kingsford.

For edge-disjoint equal-length instances, which include duplicate-merged \CSA{} instances, we give a $3/2$-approximation. When every pointer has at most two sources, we improve the ratio to $1.12585$ using MAX 2-AND. We also introduce the \emph{rainbow-path length} $\lambda$. If $\lambda\le2$, the problem can be solved exactly in polynomial time. For bounded $\lambda$, the matching algorithm gives approximation ratio $(3\lambda+1)/(2(\lambda+1))$, while for $\lambda\le3$ a reduction to 2--3-Set Packing improves the ratio to $9/8+\varepsilon$.

We also study a question specific to the complete-occurrence model: given abstract source sets, can they actually be the complete occurrence sets of strings in one common reference? We show that this realizability problem is polynomial-time solvable over an unrestricted alphabet in the explicit-position model.

Finally, we give several exact parameterized algorithms. With $M=\sum_i|S_i|$, we obtain running time $O(2^rM)$ for $r$ shared boundaries, $O(q^k\poly(N))$ for boundary budget $k$ and maximum source multiplicity $q$, $O(2^aM)$ for source ambiguity $a=\sum_i(|S_i|-1)$, and $O(q^{t+1}\poly(N))$ for pointer-interaction treewidth $t$. For edge-disjoint equal-length instances, we also give an $O(2^wM)$ algorithm for occurrence-span width $w$. In addition, \CSA{} is fixed-parameter tractable in the alphabet size $\sigma$ and the maximum pointer length $L$.

\section{Preliminaries}

Let $\Sigma$ be a finite alphabet and let $R=R[1]\cdots R[n]\in\Sigma^n$ be a reference string. For a string $P\in\Sigma^\ell$, let $\Occ_R(P)=\{a\in\{0,\ldots,n-\ell\}:R[a+1\ldots a+\ell]=P\}$ be the set of starting positions of all occurrences of $P$ in $R$.

We use the positions $0,1,\ldots,n$ between consecutive characters of $R$ as \emph{reference boundaries}. An occurrence of a string of length $\ell$ starting at position $a$ defines a \emph{source} $s=(a,\ell)$ with boundary set $\bd(s)=\{a,a+\ell\}$.

\subsection{Explicit and Complete Source Assignment}

A pointer is an identity $i$ together with a string $P_i$. Different pointers may represent the same string.

\begin{problem}[\ESA]\label{prob:esa}
Given a reference string $R$, pointer strings $P_1,\ldots,P_p$, and for every pointer $i$ a nonempty set $S_i$ of admissible sources, where every source in $S_i$ corresponds to an occurrence of $P_i$ in $R$, choose one source
$s_i\in S_i$ for each pointer, forming a feasible
source assignment $F=(s_1,\ldots,s_p)$, so as to
minimize   $\operatorname{cost}(F)=\left|\bigcup_{i=1}^p\bd(s_i)\right|$.
\end{problem}

In \ESA{}, the admissible sources are listed explicitly and need not contain every occurrence of the pointer string.

\begin{problem}[\CSA]\label{prob:csa}
Given a reference string $R$ and nonempty pointer strings $P_1,\ldots,P_p$, each occurring at least once in $R$, let $S_i$ contain exactly all occurrences of $P_i$ in $R$. Choose one source $s_i\in S_i$ for every pointer so as to minimize $\left|\bigcup_{i=1}^p\bd(s_i)\right|$.
\end{problem}

Thus, \CSA{} is the restriction of \ESA{} in which no occurrence of a pointer string may be omitted from its source set.

\begin{example}
Let $R=\texttt{abcabc}$, $P_1=\texttt{abc}$, and $P_2=\texttt{bc}$. Then $\Occ_R(P_1)=\{0,3\}$ and $\Occ_R(P_2)=\{1,4\}$. In \CSA{}, all four occurrences are available. Choosing the sources $(0,3)$ for $P_1$ and $(1,2)$ for $P_2$ uses the boundary sets $\{0,3\}$ and $\{1,3\}$, and therefore has cost $3$. Hence the optimum \CSA{} cost is $3$.

Now consider an \ESA{} instance with the same reference and pointer strings, but with $S_1=\{(0,3)\}$ and $S_2=\{(4,2)\}$. The only possible assignment uses the boundary sets $\{0,3\}$ and $\{4,6\}$, which are disjoint, so its cost is $4$.
\end{example}

If two pointers in a \CSA{} instance represent the same string, then they have the same source set. There is therefore an optimal solution that assigns them the same source, so we merge duplicate pointer strings in \CSA{}. This is not valid in \ESA{}, where equal pointer strings may have different explicitly given source sets.

\subsection{Structural notation and graph view}
\label{sec:structural-notation}

Let $p$ be the number of pointers, let $M=\sum_i |S_i|$ be the total number of
sources, and let $q=\max_i |S_i|$ be the maximum number of sources of one pointer.
We define the \emph{source ambiguity} as $a=\sum_i(|S_i|-1)=M-p$. Thus $a$
counts the total number of source choices beyond one source per pointer.

A reference boundary is \emph{shared} if it belongs to a source of at least two
different pointers. Let $C$ be the set of shared boundaries and let $r=|C|$.
A boundary outside $C$ is called \emph{private}. Such a boundary can occur only
in sources of one pointer.

We consider graphs associated with Source Assignment
instances and their feasible solutions. The \emph{source graph} $G_S$ is an edge-colored
multigraph whose vertices are the reference
boundaries $0,\ldots,n$. For every pointer $i$
and every source $s=(a,\ell)\in S_i$, we add
an edge with endpoints $a$ and $a+\ell$ and
assign it color $i$.

In \ESA{}, two sources belonging to different
pointers may have the same boundary pair,
resulting in parallel edges of different colors.
However, after merging duplicate pointer strings
in \CSA{}, no such parallel edges remain, and
$G_S$ is a simple edge-colored graph.

Thus, sources correspond to edges and pointers
to colors. A feasible source assignment selects
exactly one edge of each color, and its cost is
the number of vertices incident with the
selected edges. We ignore isolated vertices,
since they do not affect the problem.

Two sources are \emph{adjacent} if they share a reference boundary. Equivalently,
their edges in $G_S$ have a common endpoint.

We call a source system \emph{edge-disjoint} if the same physical source interval
does not belong to two different pointers. Thus there are no distinct pointers
$i$ and $j$ and sources $s\in S_i$ and $s'\in S_j$ with
$\bd(s)=\bd(s')$. Edge-disjoint sources may still be adjacent: two different
source intervals may share one boundary.

\begin{observation}\label{obs:csa-edge-disjoint}
After duplicate pointer strings are merged, every \CSA{} instance is
edge-disjoint.
\end{observation}


The \emph{pointer-interaction graph} $H_I$ has one vertex for every pointer.
Two different pointers $i$ and $j$ are adjacent in $H_I$ if some source in
$S_i$ and some source in $S_j$ share a reference boundary. When needed by an
algorithm, we store one such pair of sources as a witness for the edge $ij$.
Let $t=\operatorname{tw}(H_I)$ denote the treewidth of $H_I$.

For a feasible assignment $F$, let $G_F$ be the subgraph of $G_S$ formed by
the selected source edges and their endpoints. We also define the
\emph{selected-adjacency graph} $H_F$ on the pointer vertices. Two pointers
$i$ and $j$ are adjacent in $H_F$ exactly when the sources selected for them
by $F$ share a reference boundary. Hence $H_F$ is always a spanning subgraph of $H_I$.
Let $A(F)=|E(H_F)|$ be the number of adjacent pairs of selected sources.

We next describe the additional structure that appears when all pointers have
the same length $L$ and the source system is edge-disjoint. Every source then
has the form $(a,L)$ and corresponds to the interval between boundaries $a$
and $a+L$. Two such sources can be adjacent only when one starts where the
other ends. Therefore, along each residue class modulo $L$, the source
intervals appear in their natural left-to-right order. Since the source system
is edge-disjoint, every boundary is incident with at most one source ending
there and at most one source starting there. Consequently, every connected
component of $G_S$ is a path.

This path structure also has a direct interpretation in the reference string.
Consider sources $s_1=(a,L),s_2=(a+L,L),\ldots,s_k=(a+(k-1)L,L)$. These
sources occupy consecutive blocks of the reference: the right boundary of each
source is the left boundary of the next one. We call such a sequence a
\emph{rainbow path} if the sources belong to pairwise distinct pointers. Thus a
rainbow path is a sequence of consecutive source occurrences in the reference string
in which no pointer appears twice.

In the source graph, the equivalent of a rainbow path is exactly a nonempty subpath of a
component of $G_S$ whose edge colors are pairwise distinct. We define the
\emph{rainbow-path length} $\lambda$ as the maximum number of sources in a
rainbow path, equivalently, the maximum number of edges in such a subpath of
$G_S$. Given the source paths in their left-to-right order, $\lambda$ can be
computed in $O(M)$ time by scanning the color sequences with a sliding window
that contains no repeated color.

For a feasible assignment $F$, every nonempty connected component of $G_F$ is
called a \emph{selected component}. In the equal-length edge-disjoint setting,
each selected component is a path. Moreover, it is a rainbow path, because a
feasible assignment chooses only one source for every pointer. Hence every
selected component contains at most $\lambda$ edges.

Let $c(F)$ be the number of selected components. Since the assignment contains
exactly $p$ selected edges and these edges form a disjoint union of paths, the
selected graph uses $p+c(F)$ vertices and contains $p-c(F)$ adjacent edge
pairs. Therefore $A(F)=p-c(F)$ and
$\operatorname{cost}(F)=p+c(F)=2p-A(F)$. Thus, in this setting, minimizing the
number of selected boundaries is equivalent to maximizing the number of
adjacencies between selected sources.

We also use an ordering of the source edges for one of our dynamic programs.
For every path component of $G_S$, order its edges from left to right, and
concatenate these orders to obtain $e_1,\ldots,e_M$. For pointer $i$, let
$f_i$ and $g_i$ be the first and last positions of its sources in this order.
Pointer $i$ is \emph{active} after position $j$ if $f_i\le j<g_i$. The
\emph{occurrence-span width} is
$w=\max_{0\le j\le M}|\{i:f_i\le j<g_i\}|$. Thus $w$ measures how many
pointer choices remain unresolved across one cut of the ordered source edges.

Finally, the source graph gives two useful equivalent formulations of \ESA{}.
Let $B_S$ be the set of all reference boundaries appearing in at least one
source. A set $X\subseteq B_S$ is feasible exactly when, for every pointer $i$,
there is a source $s\in S_i$ with $\bd(s)\subseteq X$. Therefore \ESA{} is a
cardinality \SCP{} instance in which the objects are reference boundaries and
the allowed pairs for pointer $i$ are the boundary pairs of its sources.
Equivalently, \ESA{} is a unit-weight \MWMCSP{} instance on $G_S$: the pointers
are the colors, and the goal is to choose a minimum number of vertices that
induce at least one edge of every color.

\section{Reductions and Hardness}\label{sec:hardness}

\subsection{Explicit Source Assignment: hardness and approximation}
\label{sec:esa-hardness}

We first consider the explicit-source model. The following reduction from Set Cover
is value-preserving up to a factor of two and therefore also preserves approximation
ratios.

\begin{theorem}\label{thm:esa-set-cover}
For every fixed $\varepsilon>0$, \ESA{} cannot be approximated in polynomial time
within a factor $(1-\varepsilon)\ln p$, unless $\mathrm{P}=\mathrm{NP}$, where $p$
is the number of pointers. This holds even over the binary alphabet when every
pointer has length one and all pointers represent the same one-character string.
In particular, \ESA{} is APX-hard under these restrictions.
\end{theorem}

\begin{proof}
We reduce from Set Cover. Let the Set Cover instance have universe
$U=\{u_1,\ldots,u_p\}$ and sets $A_1,\ldots,A_m\subseteq U$, and assume that
every element belongs to at least one set. Define $R=(\texttt{10})^m$. The
character \texttt{1} occurs exactly at starts $0,2,\ldots,2m-2$.

For every element $u_i$, create a pointer $i$ representing the one-character
string $P_i=\texttt{1}$. Its admissible sources are
$S_i=\{(2(j-1),1):u_i\in A_j\}$. Thus source $(2(j-1),1)$ represents choosing
set $A_j$ for element $u_i$.

The source corresponding to $A_j$ has boundary pair
$\{2(j-1),2(j-1)+1\}$. These boundary pairs are disjoint for different values
of $j$. Hence, if a source assignment uses $k$ distinct source positions, its
cost is exactly $2k$.

A Set Cover consisting of $k$ sets gives a source assignment of cost at most
$2k$: assign each pointer to one selected set containing its corresponding
universe element. Conversely, if a source assignment uses $k$ distinct source
positions, the corresponding $k$ sets cover $U$. Therefore
$\OPT_{\ESA}=2\OPT_{\mathrm{SetCover}}$.

More generally, from any \ESA{} solution of cost $C$ we obtain a Set Cover of
size $C/2$. Thus an $\alpha$-approximation for these \ESA{} instances would give
an $\alpha$-approximation for Set Cover. The $(1-\varepsilon)\ln p$
inapproximability of Set Cover~\cite{DinurSteurer2014} therefore transfers
directly to \ESA{}.
\end{proof}

The graph formulation also transfers known positive approximation results to
\ESA{}. Recall that the source graph has one color for every pointer and one edge
for every admissible source. Selecting a minimum number of reference boundaries
that induce an edge of every color is exactly the unit-weight
\MWMCSP{} problem.

\begin{proposition}\label{prop:esa-general-approx}
Let $q=\max_i|S_i|$. General \ESA{}, with arbitrary pointer lengths, admits a
polynomial-time $O(\sqrt{q\log p})$-approximation.
\end{proposition}

\begin{proof}
Starting from the source graph $G_S$, we merge
parallel edges having the same endpoints and
assign to the resulting edge all their colors.
The resulting graph is an instance of unit-weight
\MWMCSP{}, where each color represents a pointer.

This transformation preserves feasible solutions
and their costs. Indeed, selecting one source for
each pointer corresponds to selecting an edge of
every color, and the reference boundaries used
by the sources are exactly the vertices incident
with the selected edges. Conversely, any vertex
set inducing at least one edge of every color
determines a feasible source assignment of no
greater cost.

The resulting graph has $p$ colors, and the
color class corresponding to pointer $i$
contains exactly $|S_i|$ edges. Hence the
maximum color-class size is $q$.

Hajiaghayi et al.~\cite{HajiaghayiEtAl2006}
give an $O(\sqrt{m\log n})$-approximation for
\MWMCSP{}, where $n$ is the number of colors
and $m$ is the maximum color-class size.
Taking $n=p$ and $m=q$ gives the claimed
$O(\sqrt{q\log p})$-approximation for \ESA{}.
\end{proof}

\subsection{Complete Source Assignment: reduction from Minimum Rainbow Subgraph}

Our main hardness result starts from a restricted form of Minimum Rainbow Subgraph. An instance consists of an edge-colored graph and asks for one edge of every color while minimizing the number of vertices incident with the selected edges. A path is \emph{properly edge-colored} if every two consecutive edges have different colors.

\begin{theorem}[H\"uffner et al.~\cite{HueffnerEtAl2015}]\label{thm:mrs-path-hard}
Minimum Rainbow Subgraph is APX-hard on properly edge-colored paths in which every color occurs at most twice.
\end{theorem}

\begin{theorem}\label{thm:hardness-large-alphabet}
\CSA{} is APX-hard even when all pointers have the same length and every pointer has at most two sources.
\end{theorem}

\begin{proof}
Let $v_0,e_1,v_1,e_2,\ldots,e_m,v_m$ be a properly edge-colored path from Minimum Rainbow Subgraph, and let $c_j$ be the color of edge $e_j$. For every color $c$, introduce one alphabet symbol $\sigma_c$ and define $R=\sigma_{c_1}\sigma_{c_2}\cdots\sigma_{c_m}$. For every color $c$, create one pointer string $P_c=\sigma_c$. Because $P_c$ has length one, its complete occurrence set is $\Occ_R(P_c)=\{j-1:c_j=c\}$. Thus the sources of pointer $c$ are exactly the path edges carrying color $c$: the occurrence at start $j-1$ has boundary pair $\{j-1,j\}$. No additional sources exist.

Choosing an occurrence of $P_c$ is therefore equivalent to choosing an edge of color $c$. Under the identification of path vertex $v_j$ with reference boundary $j$, the vertices incident to all chosen path edges are exactly the selected source boundaries. Hence feasible solutions correspond bijectively and preserve objective value.

The reduction is polynomial and optimum-preserving. Every pointer has length one, and because every color in the source instance occurs at most twice, every pointer has at most two sources. APX-hardness follows from \cref{thm:mrs-path-hard}.
\end{proof}

The isolated-source pruning rule of Qiu and Kingsford removes a nonunique source when neither of its two boundaries is shared with a source of another pointer. Such a source cannot create boundary sharing and is therefore never better than another available source of the same pointer. In our path construction, every source shares a boundary with a neighboring source of a different pointer because the path is properly edge-colored. The same is true for the two terminal edges when the path has at least two edges. Hence the pruning rule removes no source from the constructed instances.

\begin{corollary}\label{cor:pruning-hard}
The APX-hardness of \cref{thm:hardness-large-alphabet} holds even after exhaustive isolated-source pruning.
\end{corollary}

\subsection{Binary alphabet}\label{sec:binary}

We next show that the problem remains APX-hard over a binary alphabet.

\begin{theorem}\label{thm:binary-hardness}
\CSA{} is APX-hard over the binary alphabet even when all pointers have one common length and every pointer has at most two sources.
\end{theorem}

\begin{proof}
Let the alphabet of the instance from \cref{thm:hardness-large-alphabet} be $\Sigma=\{\sigma_1,\ldots,\sigma_s\}$ and let $h=\lceil\log_2 s\rceil$. Give every symbol a distinct $h$-bit identifier. Encode a bit by $\phi(0)=10$ and $\phi(1)=11$. For an identifier $b_1\cdots b_h$, define $H(\sigma)=\phi(b_1)\cdots\phi(b_h)1$ and $C(\sigma)=0001H(\sigma)$. Every codeword has length $L=2h+5$.

Inside $H(\sigma)$ there are no two consecutive zeros and the last bit is one. Therefore the marker $0001$ appears only at the start of a codeword. Every occurrence of an encoded pointer $C(P)$ is thus aligned with codeword boundaries. Since the code has fixed length and is injective, occurrences of $P$ in $R$ are in bijection with occurrences of $C(P)$ in $C(R)$.

An original boundary $b$ becomes the encoded boundary $Lb$. This map is injective, so it preserves equality of boundaries, boundary sharing, costs, and the number of sources of every pointer. Applying the encoding to the instances from \cref{thm:hardness-large-alphabet} gives a polynomial and optimum-preserving reduction to binary \CSA{}. All original pointers have length one, so all encoded pointers have the same length $L$, and every pointer still has at most two sources.
\end{proof}

For an alphabet of size $\sigma\ge2$, a duplicate-merged instance with $p$ distinct pointers of one common length $L$ satisfies $p\le\sigma^L$, and therefore $L\ge\lceil\log_\sigma p\rceil$. Thus the $\Theta(\log p)$ common length used by the binary reduction is asymptotically optimal.

\section{Approximation Algorithms}\label{sec:approximation-algorithms}

Throughout this section, we consider edge-disjoint equal-length source systems and use the notation of \cref{sec:structural-notation}. In particular, $G_S$ is a disjoint union of paths, and for every feasible assignment $F$ we have $A(F)=p-c(F)$ and $\operatorname{cost}(F)=2p-A(F)$.

\subsection{A $3/2$-Approximation via Maximum Matching}
\label{sec:approx-matching}

We first allow an arbitrary number of sources per pointer. Recall the pointer-interaction graph $H_I$ from \cref{sec:structural-notation}. In the present equal-length setting, an edge $ij$ of $H_I$ means that pointers $i$ and $j$ have sources that are consecutive on a source path and can therefore share one boundary. For every edge of $H_I$, we keep one such pair of sources as a witness. This is the auxiliary-graph approach used for Minimum Rainbow Subgraph~\cite{CamachoSchiermeyerTuza2010}.

\begin{theorem}\label{thm:three-halves}
Edge-disjoint equal-length \ESA{} admits a polynomial-time $3/2$-approximation, with no restriction on the number of sources per pointer.
\end{theorem}

\begin{proof}
We construct a feasible source assignment as follows. Compute a maximum matching $D$ of $H_I$. For every matched edge $ij\in D$, assign pointers $i$ and $j$ to the stored witnessing sources. Assign every unmatched pointer any admissible source.

Because $D$ is a matching, no pointer is assigned twice. Each matched pair gives at least one adjacency between selected sources, so the returned solution has cost at most $2p-|D|$.

Let $F^\star$ be an optimal assignment and let $A^\star=A(F^\star)$. The graph $H_{F^\star}$ is a spanning subgraph of $H_I$. Since $G_{F^\star}$ is a disjoint union of paths, $H_{F^\star}$ is also a disjoint union of paths with $A^\star$ edges. Every path forest has a matching of size at least $A^\star/2$, and every such matching edge is present in $H_I$. Therefore $|D|\ge A^\star/2$.

Since $\operatorname{OPT}=2p-A^\star$, we obtain $\operatorname{ALG}/\operatorname{OPT}\le(2p-A^\star/2)/(2p-A^\star)\le3/2$, where the last inequality follows from $0\le A^\star\le p$. Constructing $H_I$ and computing a maximum matching can both be done in polynomial time.
\end{proof}

\subsection{Bounded rainbow-path length}\label{sec:rainbow-path}

We now study how the approximation improves when the rainbow-path length $\lambda$ is small. Recall that every selected component contains at most $\lambda$ edges and that $A(F)=p-c(F)$.

\begin{theorem}\label{thm:rainbow-matching}
If $\lambda\le2$, edge-disjoint equal-length \ESA{} is solvable exactly in polynomial time. For $\lambda\ge3$, the algorithm of \cref{thm:three-halves} has approximation ratio at most $(3\lambda+1)/(2(\lambda+1))$.
\end{theorem}

\begin{proof}
Suppose first that $\lambda\le2$. Every selected component contains at most two edges, so $H_F$ is a matching for every feasible assignment $F$. In particular, for an optimal assignment $F^\star$, the graph $H_{F^\star}$ is a matching in $H_I$ of size $A^\star=A(F^\star)$. Let $D$ be a maximum matching in $H_I$. Then $|D|\ge A^\star$. Selecting the witnessing sources for the edges of $D$ and an arbitrary source for every unmatched pointer gives a feasible assignment of cost at most $2p-|D|\le2p-A^\star=\operatorname{OPT}$. Hence the assignment is optimal.

Now suppose that $\lambda\ge3$. Let $c^\star=c(F^\star)$. Every selected component has at most $\lambda$ edges and the selected components contain exactly $p$ edges in total, so $c^\star\ge p/\lambda$. Since $A^\star=p-c^\star$, we obtain $A^\star\le p(1-1/\lambda)$. By \cref{thm:three-halves}, $\operatorname{ALG}/\operatorname{OPT}\le(2p-A^\star/2)/(2p-A^\star)$. The right-hand side is increasing in $A^\star$, and substituting the preceding upper bound gives $\operatorname{ALG}/\operatorname{OPT}\le(3\lambda+1)/(2(\lambda+1))$.
\end{proof}

For $\lambda\le3$, we can obtain a better approximation. Recall that in 2--3-Set Packing, every set has size two or three: sets of size two have weight one, while sets of size three have weight two. The goal is to choose pairwise disjoint sets of maximum total weight. Neuwohner~\cite{Neuwohner2023} gives a $(4/3+\varepsilon)$-approximation for this problem.

\begin{theorem}\label{thm:rainbow-setpacking}
For every fixed $\varepsilon>0$, edge-disjoint equal-length \ESA{} with $\lambda\le3$ admits a polynomial-time $(9/8+\varepsilon)$-approximation, with no restriction on source multiplicity.
\end{theorem}

\begin{proof}
Build a weighted 2--3-Set Packing instance whose ground set consists of the pointer identities. For every pair of adjacent source edges in $G_S$ with distinct colors, create a set containing their two pointer identities and give it weight one. For every three-edge subpath of $G_S$ with pairwise distinct colors, create a set containing its three pointer identities and give it weight two. For each set, store the corresponding source edges as a witness; if the same set is generated more than once, keep one witness.

Since sets in a packing are pairwise disjoint, their pointer identities are distinct. Hence the source choices represented by their witnesses are compatible and can be extended to a feasible assignment by choosing arbitrary sources for all remaining pointers. The resulting assignment realizes at least the total packing weight in adjacencies.

Conversely, in an optimal assignment $F^\star$, every selected component contains at most three edges. Represent every selected component with two or three edges by its corresponding set. These sets are pairwise disjoint and their total weight is $A^\star$. Thus the optimum packing weight is exactly $A^\star$.

Let $\rho=4/3+\delta$ be the approximation ratio used for 2--3-Set Packing. The resulting source assignment has at least $A^\star/\rho$ adjacencies. Since every selected component in $F^\star$ has at most three edges, $A^\star\le2p/3$. Therefore $\operatorname{ALG}/\operatorname{OPT}\le(2p-A^\star/\rho)/(2p-A^\star)\le3/2-1/(2\rho)$. As $\delta$ tends to zero, this bound tends to $9/8$. Choosing $\delta$ sufficiently small as a function of $\varepsilon$ gives the claimed ratio.
\end{proof}

\subsection{A $1.12585$-Approximation via MAX 2-AND}\label{sec:approx}

We now additionally assume that every pointer has at most two sources. A pointer with two sources can be represented by one Boolean variable. We use MAX 2-AND to encode which pairs of selected sources are adjacent. In unweighted MAX 2-AND, each constraint is a conjunction of two Boolean literals, and the goal is to maximize the number of satisfied constraints. Repeated constraints are allowed.

\begin{theorem}\label{thm:approx}
Edge-disjoint equal-length \ESA{} with at most two sources per pointer admits a $(2-\alpha_{\mathrm{2AND}})$-approximation, where $\alpha_{\mathrm{2AND}}$ is the approximation ratio for unweighted MAX 2-AND. Using $\alpha_{\mathrm{2AND}}\ge0.87415$~\cite{BrakensiekEtAl2026}, the ratio is at most $1.12585$.
\end{theorem}

\begin{proof}
For every pointer $i$ with two sources $s_i^0$ and $s_i^1$, introduce a Boolean variable $x_i$, where $x_i=b$ means that source $s_i^b$ is selected. A pointer with only one source needs no variable.

Consider every pair of adjacent source edges belonging to different pointers. If both pointers have two sources, say the adjacent sources are $s_i^b$ and $s_j^{b'}$, add the MAX 2-AND constraint $[x_i=b]\wedge[x_j=b']$, where $[x_i=1]$ denotes $x_i$ and $[x_i=0]$ denotes $\neg x_i$.

Suppose instead that exactly one of the two pointers has two sources. Let $\ell$ be the literal selecting the source that participates in the adjacency. Introduce a new variable $y$ and add the two constraints $\ell\wedge y$ and $\ell\wedge\neg y$. If $\ell$ is true, exactly one of these constraints is satisfied, independently of the value of $y$; if $\ell$ is false, neither is satisfied. Thus this pair contributes exactly one satisfied constraint exactly when the corresponding source adjacency is selected.

If both pointers have one source, their adjacency is unavoidable. Let $C_0$ be the number of such adjacencies. We do not add constraints for them because they contribute the same amount to every source assignment.

For every source assignment $F$, the corresponding Boolean assignment therefore satisfies exactly $A(F)-C_0$ constraints. Conversely, every Boolean assignment determines a source assignment with the same relation. The auxiliary variables do not affect this correspondence because each unary gadget contributes either zero or one independently of the value chosen for its new variable.

Let $A^\star$ be the maximum possible number of selected adjacencies. The optimum value of the MAX 2-AND instance is $A^\star-C_0$. An $\alpha_{\mathrm{2AND}}$-approximation therefore returns a source assignment $F$ satisfying $A(F)\ge C_0+\alpha_{\mathrm{2AND}}(A^\star-C_0)\ge\alpha_{\mathrm{2AND}}A^\star$.

Using $\operatorname{cost}(F)=2p-A(F)$, we obtain $\operatorname{ALG}/\operatorname{OPT}\le(2p-\alpha_{\mathrm{2AND}}A^\star)/(2p-A^\star)$. Since $0\le A^\star\le p$, this ratio is at most $2-\alpha_{\mathrm{2AND}}$. Using $\alpha_{\mathrm{2AND}}\ge0.87415$ gives $1.12585$. The reduction and decoding are polynomial-time.
\end{proof}

\section{Recognizing Complete String-Realizable Source Systems}
\label{sec:realizability}

The graph view does not capture one important condition of \CSA{}: every source set must contain all occurrences of one string in the same reference. We therefore ask when an abstract family of source sets can be realized in this way.

An instance gives a reference length $n$ and, for each pointer $i$, a positive length $\ell_i$ and a nonempty set $T_i\subseteq\{0,\ldots,n-\ell_i\}$ of desired occurrence starts. We use the explicit-position model, where the $n$ reference positions are part of the input; equivalently, $n$ may be viewed as unary encoded.

\begin{problem}[Source-Set Realizability]\label{prob:realizability}
Given $n$ and pairs $(\ell_i,T_i)$, decide whether there is a reference string $R$ over some alphabet and, for every pointer $i$, a string $P_i$ of length $\ell_i$ such that $\Occ_R(P_i)=T_i$.
\end{problem}

For each pointer $i$, choose one anchor start $a_i\in T_i$. Every required occurrence at $s\in T_i$ forces $R[a_i+d]=R[s+d]$ for every offset $0\le d<\ell_i$. Let $\equiv$ be the equivalence relation on reference positions generated by all these forced equalities.

\begin{theorem}\label{thm:realizability}
In the explicit-position model, Source-Set Realizability is polynomial-time solvable. The source system is realizable if and only if, for every pointer $i$ and every unwanted start $x\notin T_i$, there is an offset $0\le d<\ell_i$ such that $x+d\not\equiv a_i+d$.
\end{theorem}

\begin{proof}
Every equality used to build $\equiv$ is forced in every realization, because all starts in $T_i$ must contain the same length-$\ell_i$ string as the anchor. Thus, if an unwanted start $x\notin T_i$ satisfies $x+d\equiv a_i+d$ for every offset $d$, then every realization also has $R[x+d]=R[a_i+d]$ for every $d$. Hence $x$ would be an unavoidable extra occurrence, so the source system is not realizable.

For the other direction, give a different alphabet symbol to every equivalence class of $\equiv$, and call the resulting string $R^\star$. Every required start $s\in T_i$ is an occurrence because $s+d\equiv a_i+d$ for every offset. Every unwanted start fails at some offset by the condition of the theorem, so it is not an occurrence in $R^\star$. Therefore $R^\star$ realizes exactly the prescribed source sets.

The equivalence classes can be computed with union--find using $O(\sum_i|T_i|\ell_i)$ union operations, and all unwanted starts can be checked in $O(\sum_i n\ell_i)$ time. This is polynomial in the explicit-position model. If $n$ is given only in binary and the positions are not explicitly represented, this bound is only pseudopolynomial in the bit length of $n$.
\end{proof}

\section{Exact and Parameterized Algorithms}\label{sec:exact-algorithms}

We now turn to exact algorithms. Let $N$ denote the size of the explicitly represented input. The parameters $M$, $q$, $a$, $r$, $t$, and $w$ are defined in \cref{sec:structural-notation}.

\subsection{Parameterization by Shared Boundaries}\label{sec:shared}

Recall that $C$ is the set of shared boundaries and $r=|C|$. Every boundary outside $C$ is private, so once we decide which shared boundaries may be used, the private contribution of each pointer can be minimized independently.

\begin{theorem}\label{thm:shared-fpt}
\ESA{} can be solved in $O(2^rM)$ time and polynomial space.
\end{theorem}

\begin{proof}
For every subset $X\subseteq C$, consider only sources whose shared endpoints are contained in $X$. For pointer $i$, define $c_i(X)=\min\{|\bd(s)\setminus C|:s\in S_i,\ \bd(s)\cap C\subseteq X\}$, where $c_i(X)=+\infty$ if no such source exists. Thus $c_i(X)$ is the minimum number of private boundaries needed by pointer $i$ when only shared boundaries from $X$ may be used.

For a fixed $X$, choose for every pointer a source attaining $c_i(X)$. Private boundaries belonging to different pointers are disjoint, so the resulting assignment has cost at most $|X|+\sum_i c_i(X)$. Conversely, let $F^\star$ be an optimal assignment and let $X^\star$ be exactly the shared boundaries used by $F^\star$. Every source selected by $F^\star$ is allowed when evaluating $X^\star$, so $|X^\star|+\sum_i c_i(X^\star)\le\operatorname{OPT}$. It follows that $\operatorname{OPT}=\min_{X\subseteq C}(|X|+\sum_i c_i(X))$.

There are $2^r$ subsets $X$. For each one, all values $c_i(X)$ can be computed by scanning the $M$ sources. The total running time is $O(2^rM)$ and only polynomial space is required.
\end{proof}

\subsection{Boundary Budget and Source Multiplicity}\label{sec:q}

Let $k$ be the maximum allowed number of selected boundaries. A pointer is satisfied by a current boundary set $X$ if some source $s\in S_i$ satisfies $\bd(s)\subseteq X$. If a pointer is not satisfied, every feasible completion must include both endpoints of one of its sources.

\begin{theorem}\label{thm:qk-fpt}
The decision version of \ESA{} with boundary budget $k$ can be solved in $O(q^k\poly(N))$ time. Consequently, \ESA{} is fixed-parameter tractable parameterized by $k+q$.
\end{theorem}

\begin{proof}
Start with $X=\emptyset$. If $|X|>k$, reject the current branch. If every pointer is satisfied, accept. Otherwise choose an unsatisfied pointer $i$. For every source $s\in S_i$, create one branch in which both boundaries in $\bd(s)$ are added to $X$.

The branching is complete because every feasible solution extending $X$ must select one source of the chosen unsatisfied pointer and therefore contains both endpoints of at least one source in $S_i$. Since the pointer is currently unsatisfied, every branch adds at least one new boundary. Thus every root-to-leaf path has at most $k$ branching levels. There are at most $q$ branches at each level, and checking whether all pointers are satisfied takes polynomial time. The running time is therefore $O(q^k\poly(N))$.
\end{proof}

\subsection{A Reduction to Vertex Cover when $q\le2$}
\label{sec:q2-vertex-cover}

When every pointer has at most two sources, the choice of reference boundaries can be written as a monotone Boolean formula with clauses of size at most two. After forced boundaries are processed, the remaining formula is exactly a Vertex Cover instance.

\begin{proposition}\label{prop:q2-vc}
After exhaustive processing of forced boundaries, \ESA{} with at most two distinct sources per pointer admits a polynomial-time optimum-preserving reduction to Minimum Vertex Cover, up to the additive cost of the forced boundaries.
\end{proposition}

\begin{proof}
Introduce one Boolean variable for every relevant reference boundary, where setting it to one means selecting that boundary. A pointer with one source having endpoints $a$ and $b$ requires both $a$ and $b$. For a two-source pointer with boundary pairs $\{a,b\}$ and $\{c,d\}$, feasibility requires $(a\wedge b)\vee(c\wedge d)$, which is equivalent to $(a\vee c)\wedge(a\vee d)\wedge(b\vee c)\wedge(b\vee d)$, including when some endpoints coincide.

Let $F$ be the set of forced boundaries. Initially, $F$ contains both endpoints of every one-source pointer. Whenever a clause becomes a unit clause, its remaining boundary must also be added to $F$. Repeating this process until no new boundary is forced leaves only binary clauses $u\vee v$ with $u,v\notin F$.

Construct a graph $G$ with one vertex for every unforced boundary occurring in a remaining clause and one edge $uv$ for every clause $u\vee v$. A set $Y$ of unforced boundaries satisfies all remaining clauses exactly when it contains at least one endpoint of every edge, that is, exactly when $Y$ is a vertex cover of $G$. Hence $\operatorname{OPT}_{\ESA}=|F|+\operatorname{OPT}_{\mathrm{VC}}(G)$. The propagation and construction are polynomial-time.
\end{proof}

\subsection{Dynamic Programming on the Pointer-Interaction Graph}
\label{sec:treewidth}

Recall the pointer-interaction graph $H_I$ and its treewidth $t$. For every relevant boundary $b$, let $P_b$ be the set of pointers having a source incident with $b$. Every two pointers in $P_b$ are adjacent in $H_I$, so $P_b$ is a clique. Therefore some bag of every tree decomposition of $H_I$ contains all of $P_b$. Assign each boundary $b$ to one such bag. For a choice of sources of the pointers in $P_b$, let $g_b$ be $1$ if at least one selected source uses $b$, and $0$ otherwise. The sum of these values over all boundaries is exactly the number of distinct selected boundaries.

\begin{theorem}\label{thm:interaction-treewidth}
Given a tree decomposition of $H_I$ of width $t$, \ESA{} can be solved in $O(q^{t+1}\poly(N))$ time. Consequently, \ESA{} is fixed-parameter tractable parameterized by $t+q$.
\end{theorem}

\begin{proof}
Root a nice tree decomposition and let $B_u$ be the bag at node $u$. A DP state at $u$ is an assignment $\eta\in\prod_{i\in B_u}S_i$, specifying one source for every pointer in the bag. Let $D_u(\eta)$ be the minimum total cost of boundaries assigned to bags in the subtree rooted at $u$, over all source choices consistent with $\eta$. Let $\operatorname{local}_u(\eta)$ be the sum of $g_b$ over all boundaries assigned to bag $u$.

For an empty leaf, the recurrence is $D_u(\emptyset)=\operatorname{local}_u(\emptyset)$. If $u$ introduces pointer $i$ and has child $v$, then $D_u(\eta)=\operatorname{local}_u(\eta)+D_v(\eta|_{B_v})$. If $u$ forgets pointer $i$ and has child $v$, then $D_u(\eta)=\operatorname{local}_u(\eta)+\min_{s\in S_i}D_v(\eta\cup\{i\mapsto s\})$. If $u$ is a join node with children $v_1$ and $v_2$ having the same bag, then $D_u(\eta)=\operatorname{local}_u(\eta)+D_{v_1}(\eta)+D_{v_2}(\eta)$. At an empty root, the optimum is $D_{\mathrm{root}}(\emptyset)$.

Every boundary is charged at exactly one bag containing all pointers that may use it. The DP therefore counts every selected boundary exactly once. The introduce, forget, and join recurrences consider all source choices while keeping consistent choices on overlapping bags, so induction over the tree decomposition proves correctness.

Every bag contains at most $t+1$ pointers and therefore has at most $q^{t+1}$ states. A nice decomposition has polynomially many nodes, so the running time is $O(q^{t+1}\poly(N))$.
\end{proof}

\subsection{Complete Occurrences: Alphabet Size and Pointer Length}
\label{sec:pointer-length}

We next use restrictions specific to \CSA{}. Let $\sigma=|\Sigma|$ be the alphabet size and let $L$ be the maximum pointer length.

\begin{theorem}\label{thm:pointer-length-fpt}
\CSA{} is fixed-parameter tractable parameterized by $\sigma+L$. In particular, for every fixed alphabet, it is fixed-parameter tractable parameterized by $L$.
\end{theorem}

\begin{proof}
After duplicate pointers are merged, the number of different pointer strings is at most $p\le\sum_{\ell=1}^{L}\sigma^\ell$. Solving \CSA{} is exactly Minimum Rainbow Subgraph on the source graph $G_S$.

Let $\Delta$ be the maximum degree of $G_S$. For a fixed boundary $b$ and a fixed length $\ell\le L$, at most one physical interval of length $\ell$ starts at $b$ and at most one ends at $b$. Since each physical interval determines one substring and duplicate pointer strings have been merged, there are at most two incident edges of length $\ell$. Hence $\Delta\le2L$.

If $\Delta\le1$, all source edges are boundary-disjoint and every feasible assignment has cost $2p$. Otherwise, apply the exact $O^*((4\Delta-4)^p)$ algorithm of H\"uffner et al.~\cite{HueffnerEtAl2015} for Minimum Rainbow Subgraph. Both $p$ and $\Delta$ are bounded by functions of $\sigma$ and $L$, so the running time is a computable function of $\sigma+L$ multiplied by $\poly(N)$. Therefore \CSA{} is FPT in $\sigma+L$.
\end{proof}

\subsection{Source Ambiguity}\label{sec:ambiguity}

Recall that the source ambiguity is $a=\sum_i(|S_i|-1)=M-p$. When every pointer has at most two sources, $a$ is exactly the number of pointers with two possible sources.

\begin{theorem}\label{thm:ambiguity}
\ESA{} can be solved in $O(2^aM)$ time and polynomial space.
\end{theorem}

\begin{proof}
Select the unique source of every one-source pointer. For every pointer with at least two sources, enumerate all of its possible source choices. The number of assignments considered is at most $\prod_{i:|S_i|\ge2}|S_i|\le\prod_{i:|S_i|\ge2}2^{|S_i|-1}=2^a$, since $x\le2^{x-1}$ for every integer $x\ge2$.

Each completed assignment can be evaluated in $O(M)$ time by marking its distinct boundary endpoints. Since every feasible source assignment appears in the enumeration, the best one is optimal. The total running time is $O(2^aM)$ and only polynomial space is required.
\end{proof}

\subsection{Dynamic Programming by Occurrence-Span Width}
\label{sec:span-width}

Recall the occurrence-span width $w$ and the edge order $e_1,\ldots,e_M$ from \cref{sec:structural-notation}. We again assume edge-disjoint equal-length sources, with no restriction on source multiplicity.

\begin{theorem}\label{thm:span-width}
Edge-disjoint equal-length \ESA{} can be solved in $O(2^wM)$ time and $O(2^w)$ additional space. Moreover, $w\le a$.
\end{theorem}

\begin{proof}
Scan the source edges in the order $e_1,\ldots,e_M$. After processing $e_1,\ldots,e_j$, a DP state contains one bit $\mu_i$ for every active pointer $i$, where $\mu_i=1$ means that a source of $i$ has already been selected, and one additional bit $z$ indicating whether the immediately preceding edge in the current source-path component was selected. At the beginning of a new path, use $z=0$.

Let $D_j(\mu,z)$ be the minimum number of selected boundaries after processing $e_1,\ldots,e_j$ among all partial assignments represented by state $(\mu,z)$. Initially, $D_0(\emptyset,0)=0$, and every other state has value $+\infty$.

Consider edge $e_j$ belonging to pointer $i$, and let $u$ be $1$ if a previous source of $i$ has already been selected and $0$ otherwise. If $u=1$, then $e_j$ must be skipped. If $u=0$ and $j=g_i$, then $e_j$ must be selected because it is the last available source of pointer $i$. Otherwise either choice is allowed. Let $b\in\{0,1\}$ indicate whether $e_j$ is selected, and let $\widehat z$ equal the previous $z$, except that $\widehat z=0$ when $e_j$ starts a new path. If $(\mu',b)$ is the resulting state, the recurrence is $D_j(\mu',b)=\min\{D_{j-1}(\mu,z)+b(2-\widehat z)\}$, where the minimum is over all predecessor states and allowed choices that produce $\mu'$. If pointer $i$ remains active, its new bit is $\mu'_i=u\vee b$; if $j=g_i$, the bit is removed.

The DP selects at most one source for every pointer because once its bit becomes one, later sources must be skipped. If no source has been selected before the last occurrence of a pointer, that last source is forced, so every pointer is selected exactly once. Since every source-path component is a path, the only previously processed source that can share a boundary with the current source is its immediate predecessor. Thus selecting the current edge contributes exactly $2-\widehat z$ new boundaries.

At every cut there are at most $w$ active pointers. Including the bit $z$, there are at most $2^{w+1}=O(2^w)$ states, and each state has at most two transitions. Only two consecutive DP layers are needed, giving running time $O(2^wM)$ and additional space $O(2^w)$.

Finally, every active pointer has at least two sources and therefore contributes at least one to $a=\sum_i(|S_i|-1)$. Hence $w\le a$.
\end{proof}

\section{Conclusions}

We separated the explicit-source model from the complete-occurrence setting suggested by string compression. The complete-occurrence restriction does not remove the difficulty of the problem: \CSA{} remains APX-hard even for binary equal-length instances with at most two sources per pointer. We also gave approximation algorithms for equal-length instances, including stronger guarantees when the source paths have small rainbow-path length, a polynomial-time test for complete source-set realizability over an unrestricted alphabet, and several exact parameterized algorithms.

Two natural directions are to improve the approximation ratio for general equal-length instances and to study which structural parameters are small on Source Assignment instances produced by real compression pipelines.

\section*{AI disclosure} The authors used GPT-5.6 Sol during the preparation of this
work to brainstorm and refine constructions, explore proof
strategies, and discuss and check mathematical arguments.
The model also assisted with proofreading and aspects of
the literature review. All mathematical claims, proofs,
and cited results were independently verified by the authors,
who take full responsibility for the accuracy and content
of the paper.

\bibliography{bibliography}
\end{document}